\documentclass{aa}

\usepackage[nomarkers,figuresonly]{endfloat}
\usepackage[]{natbib}

\begin{document} 

   \title{Dust Impact Monitor (SESAME-DIM) on board Rosetta/Philae: Millimetric particle flux at comet 67P/Churyumov-Gerasimenko}
   \titlerunning{Estimation of particle flux based on SESAME-DIM measurements}

%

   \author{Attila Hirn \inst{1}\fnmsep\inst{2}\fnmsep\thanks{\emph{contact address:} \email{attila.hirn@energia.mta.hu}}
          \and Thomas Albin\inst{3}\fnmsep\inst{4}\fnmsep\inst{2}
          \and Istv\'an Ap\'athy \inst{1}
	\and Vincenzo Della Corte \inst{5}
          \and Hans-Herbert Fischer\inst{6}
          \and Alberto Flandes\inst{7}\fnmsep\inst{2}
          \and Alexander Loose \inst{2}
          \and Attila P\'eter \inst{1}
          \and Klaus~J. Seidensticker \inst{8}
          \and Harald Kr\"uger\inst{2}
          }

   \authorrunning{A. Hirn et al.}

   \institute{Centre for Energy Research, Hungarian Academy of Sciences, Konkoly Thege Mikl\'os \'ut 29-33, 1121 Budapest, Hungary
          \and  Max-Planck-Institut f\"ur Sonnensystemforschung, Justus-von-Liebig-Weg 3, 37077 G\"ottingen, Germany
          \and Institut f\"ur Raumfahrtsysteme, University Stuttgart, Pfaffenwaldring 29, 70569 Stuttgart, Germany
          \and Medical Radiation Physics, Faculty VI, Carl von Ossietzky University, Georgstrasse 12, 26121 Oldenburg, Germany
	\and Institute for Space Astrophysics and Planetology (IAPS), National Institute for AstroPhysics (INAF),
		Via Fosso del Cavaliere 100, 00133 Roma, Italy
          \and Deutsches Zentrum f\"ur Luft- und Raumfahrt, Raumflugbetrieb und Astronautentraining, MUSC,
		Linder H\"ohe, 51147 K\"oln, Germany
          \and Ciencias Espaciales, Instituto de Geof\'isica, Universidad Nacional Aut\'onoma de M\'exico,
		Coyoac\'an 04510, M\'exico, D.F.
          \and Deutsches Zentrum f\"ur Luft- und Raumfahrt, Institut f\"ur Planetenforschung,
		Rutherfordstra{\ss}e 2, 12489 Berlin, Germany
             }

   \date{Received; accepted}

 
  \abstract
   {The Philae lander of the Rosetta mission, aimed at the in situ investigation of comet 67P/Churyumov-Gerasimenko, was deployed to the surface of the comet nucleus on 12 November 2014 at 2.99~AU heliocentric distance. The Dust Impact Monitor (DIM) as part of the Surface Electric Sounding and Acoustic Monitoring Experiment (SESAME) on the lander employed piezoelectric detectors to detect the submillimetre- and millimetre-sized dust and ice particles emitted from the nucleus.}
   {We determine the upper limit of the ambient flux of particles in the measurement range of DIM based on the measurements performed with the instrument during Philae’s descent to its nominal landing site Agilkia at distances of about 22~km, 18~km, and 5~km from the nucleus barycentre and at the final landing site Abydos.}
   {The geometric factor of the DIM sensor was calculated assuming an isotropic ambient flux of the submillimetre- and millimetre-sized particles. For the measurement intervals when no particles were detected the maximum true impact rate was calculated by assuming Poisson distribution of the impacts, and it was given as the detection limit at a 95\% confidence level. The shading by the comet environment at Abydos was estimated by simulating the pattern of illumination on Philae and consequently the topography around the lander.}
   {Based on measurements performed with DIM, the upper limit of the flux of particles in the measurement range of the instrument was of the order of $10^{-8}-10^{-7}\mathrm{m}^{-2}\mathrm{s}^{-1}\mathrm{sr}^{-1}$ during descent. The upper limit of the ambient flux of the submillimetre- and millimetre-sized dust and ice particles at Abydos was estimated to be $1.6\cdot10^{-9} \mathrm{m}^{-2}\mathrm{s}^{-1}\mathrm{sr}^{-1}$ on 13 and 14 November 2014. A correction factor of roughly $1/3$ for the field of view of the sensors  was calculated based on an analysis of the pattern of illumination on Philae.}
   {Considering particle speeds below escape velocity, the upper limit for the volume density of particles in the measurement range of DIM was constrained to $10^{-11}\,\mathrm{m}^{-3}-10^{-12}\,\mathrm{m}^{-3}$. Results of the calculations performed with the GIPSI tool on the expected particle fluxes during the descent of Philae were compatible with the non-detection of compact particles by the DIM instrument.
}    


   \maketitle

   \bibliographystyle{aa} 


\section{Introduction}

\label{sec_introduction}
 
After its more than 10-year cruise to comet 67P/Churyumov-Gerasimenko (hereafter 67P/C-G), the Rosetta spacecraft \citep{glassmeier2007} reached its target on the 6 August 2014 to start a series of in situ measurements from around the nucleus. In addition to the 11 orbiter experiments, Rosetta also carried a lander, named Philae \citep{bibring2007}, which was deployed onto the surface of the nucleus of the comet on 12 November 2014 \citep{biele2015}.

The Dust Impact Monitor (DIM) of the Surface Electric Sounding and Acoustic Monitoring Experiment (SESAME) package \citep{seidensticker2007} on board the lander was one of the instruments that were active and operating not only 
during the first few days after the landing of Philae (first science sequence -- FSS) on the nucleus surface at the final landing site Abydos, but also during the separation, descent, and landing (SDL) phase of the mission. The DIM instrument was designed to measure the flux of submillimetre- and millimetre-sized dust and ice particles emitted from the nucleus by means of $3\times3$ piezoelectric sensor segments made of PNZT7700 (Pb, Ni, Zi, Ti, hereafter referred to as PZT) and mounted on three sides of a cube. From the signal properties measured with the associated SESAME common electronics, the mass and the speed of the impacting particles could be constrained assuming given density and elastic material properties. Since the sensor sides are oriented in three mutually orthogonal directions, an assessment of the directionality of the impacting particles might be also made provided that the number of impacts is statistically sufficient \citep{seidensticker2007, flandes2013, flandes2014}.

DIM was operated during three mission phases of Philae at the comet \citep{krueger2015}: before separation, during descent, and at the final landing site. In the mission phase before Philae's separation from Rosetta, at altitudes between approximately 8 and 23~km from the nucleus surface, DIM was significantly obscured by structures of Rosetta and no particles were detected. During Philae's descent to its nominal landing site Agilkia, DIM detected one approximately millimetre-sized particle at an altitude of 2.4 km. This is the closest ever detection at a cometary nucleus by a dedicated in situ dust detector. The material properties of the detected particle are compatible with a porous particle having a bulk density of approximately $\mathrm{250\,kg\,m^{-3}}$. At Philae's final landing site, Abydos, DIM detected no particle impacts.

In this paper we present upper limits of the flux of particles in the measurement range of the DIM instrument in the two operational phases after the release of Philae. Phases preceding the release are not considered in our analysis because of the complexity of the shielding geometry and a reduced geometric factor of the shaded sensors. Measurements of the particle flux on the orbiter are provided by the Grain Impact Analyser and Dust Accumulator (GIADA) team for this period with much greater sensitivity and a different measurement range \citep{dellacorte2015, rotundi2015}.
We discuss in detail the effects of shading by the detector frame and the body of the lander on the geometric factor of the DIM sensor, and also address the effects of the local environent. A rough estimation on the upper limit of the volume density of particles in DIM's measurement range is presented.


\section{Dust Impact Monitor}

\label{sec_dust_impact_monitor}


\subsection{Detector geometry}

\label{subsec_det_geometry}

\begin{figure}
   \centering
   \includegraphics[width=\hsize]{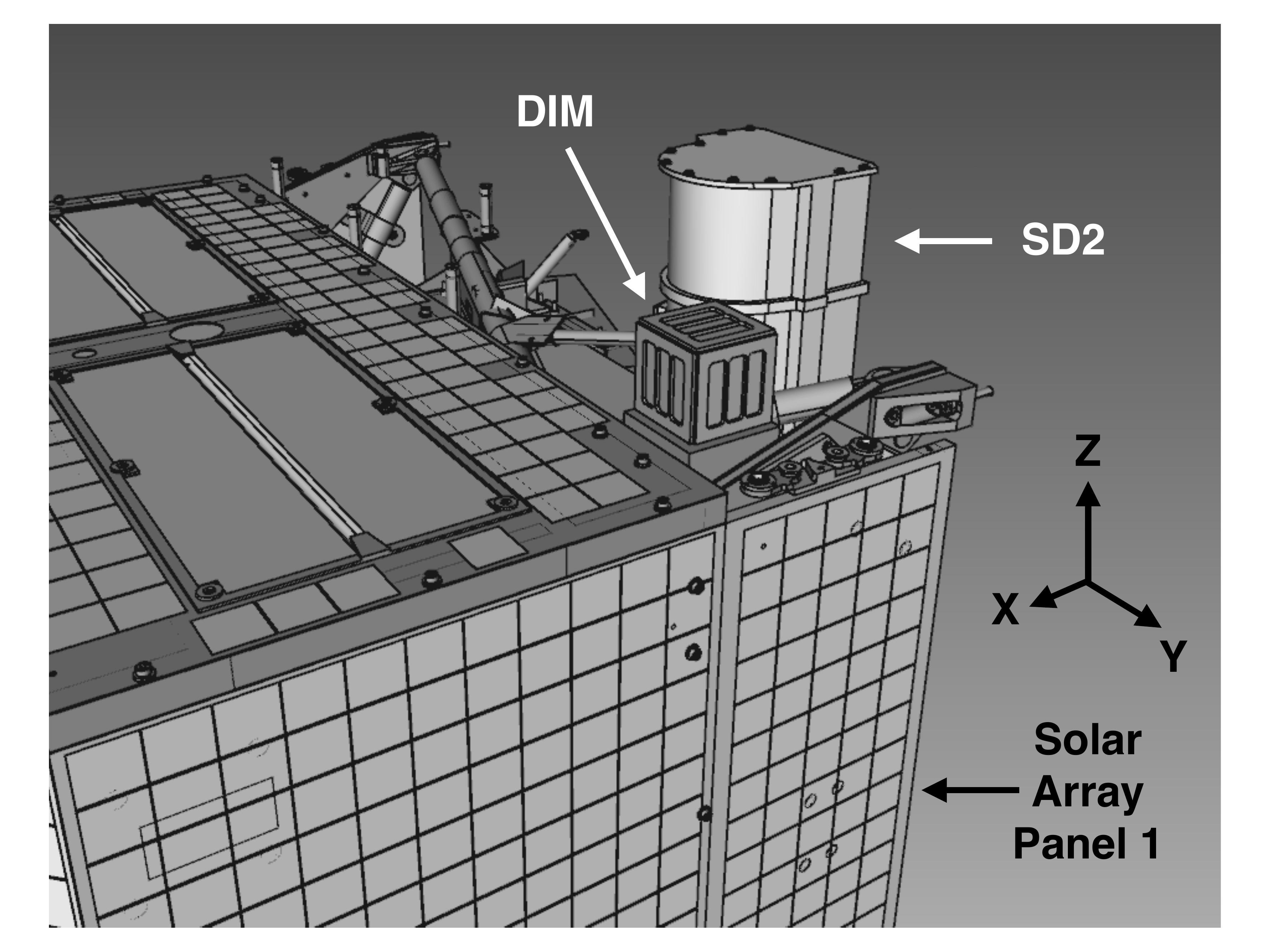}
   \caption{DIM on the top of Philae. The sensor is visible in the corner \citep{krueger2015}. Credits: ESA/ATG medialab.
          }
    \label{Fig_Philae_DIM}
\end{figure}

The DIM cube of dimensions $71.5\,\mathrm{mm}\times71.5\,\mathrm{mm}\times69.0\,\mathrm{mm}$ is mounted on the top face of the lander, above Philae's balcony, with sensor sides pointing in the +X, +Y, and +Z directions in the Philae coordinate system (Fig.~\ref{Fig_Philae_DIM}). The -X and -Y sides are covered with aluminum plates, whereas the -Z side is left open for cabling and mounting purposes. The three PZT segments on the active sides have dimensions $50.0\,\mathrm{mm}\times16.2\,\mathrm{mm}\times1.0\,\mathrm{mm}$ and they are separated by 1.5~mm (Fig.~\ref{Fig_max_angles}). They all lie 2.3~mm below the frame of the DIM cube. Impacts incident on different sensor segments belonging to a given sensor side are not distinguished by the electronics.

The PZT segments are significantly shaded by Philae's structure and by the sensor frame; the amount of shading differs from one side of the sensor to the other (see Fig.~\ref{Fig_Philae_DIM}). The field of view (FoV) of the +X side is limited mostly to the +Z direction owing to the structure of the lander. However, the +Y sensor side, being closer to the edge of the structure, is only partially shielded, mostly by Solar Array Panel 1, for particles approaching from the –Z half-space. Because it is close to the drill box of the drill, sample, and distribution (SD2) subsystem protuding approximately 150~mm above Philae's solar hood and 75~mm above DIM's Z side (see Fig.~\ref{Fig_Philae_DIM}), the +Z sensor side is also partially shielded from particles coming from the (-X; -Y; +Z) region, but that side is still the least shielded of the three active sides of DIM.

\subsection{Measurement technique}

\label{subsec_meas_technique}

When a PZT segment on DIM  is hit by a particle, the sensor generates a damped, closely sinusoidal electrical signal. Calibration measurements performed on the ground with different types of test particles impacting on the DIM sensor at different speeds \citep{peter2002, flandes2013, flandes2014} have shown that the impacts can be described and analysed using the Hertz
 theory of contact mechanics \citep{hertz1882, johnson2003}. After recording the amplitude $(U_m)$ and the width of the first half-sine pulse $(T_c)$, it is possible to constrain the radius $R$ and the impact speed $v$ of the particle \citep{seidensticker2007, flandes2013} according to Eqs. \ref{eq_ampl} and \ref{eq_tc},

\begin{equation}
   U_m=\dfrac{3.03d_{33}E^{0.4}_r\rho^{0.6}R^2v^{1.2}}{C}\,,  \label{eq_ampl}
\end{equation} 

\begin{equation}
   T_c=5.09\,\Bigg(\dfrac{R^5\rho^2}{vE^2_r}\Bigg)^{1/5}\,,  \label{eq_tc}
\end{equation} 

where $d_{33}$ is the piezoelectric stress constant of the PZT, $E_r$ the combined reduced Young's modulus of the sensor and the impinging particle, $\rho$ the density of the impacting particle, and $C$ the capacitance of the sensor plate. 

The signals coming from the sensor segments are first amplified with a logarithmic amplifier. The instrument detects a signal only if the amplified signal crosses an adaptive detection threshold voltage defined as the sum of an adjustable margin and the signal average determined by the DIM electronics with a time constant of approximately 1~s. The margin can be increased in steps of 10~dB in the range from 10~dB to 70~dB. Each step changes the threshold voltage by approximately 0.3~V. If the amplified signal crosses the detection threshold less than 1~ms after the single event measurement is initialized, the signal is classified as a ``false event'', else it is accepted as the beginning of a potential real impact. If no second threshold crossing is detected within a time interval pre-defined for the given operational mode, the event is ruled out as a ``long event''. A new single event measurement is initialized only after some waiting and latency periods adding up to a total instrument dead time of approximately 10~ms. A more detailed description of the DIM signal processing is given in \citet{peter2001, fischer2014} and \citet{ krueger2015}.
\citet{flandes2013} have shown that signals with amplitudes in the interval 0.2~mV $< U_m <$ 15~mV deliver measurement values within the expected theoretical behaviour and from this they have determined an approximated experimental range of detection radii based on Eq.~\ref{eq_ampl} \citep{krueger2015}. The intervals are reported for the different operational modes in Sects.~\ref{subsubsec_sdl_op} and \ref{subsubsec_fss_op}.


\subsection{Operation}

\label{subsec_operation}

After Rosetta's launch in March 2004, the health and the performance of the DIM instrument were regularly checked and interference tests were executed in the frame of payload check-outs performed approximately every six months until the spacecraft entered deep space hibernation in 2011. After the wake-up of Rosetta and its lander Philae in 2014, further tests were performed in order to guarantee that DIM was working properly. A detailed description of these operations (health-checks, tests, and measurements) is given in \citet{fischer2012} and \citet{krueger2015} and the corresponding SESAME Lander Instrument Operations Requests (LIOR) documents. In the present paper we focus exclusively on the measurement modes used in the SDL and FSS phases of the mission.


\subsubsection{Measurement mode during the separation, descent, and landing phase}

\label{subsubsec_sdl_op}

During the SDL phase, measurements were performed in the so-called Burst Continuous Test2 mode (BCT2). This measurement mode delivers the measured raw peak amplitude $U_m$, the impact duration $T_c$, and the time of detection for the first 350 detected events on a given sensor side. The total number of detected events, false events, and long events are also recorded. On 12 November 2014, after Philae's separation from Rosetta at 08:35:00 UTC, three measurement blocks were conducted at distances of about 22~km, 18~km, and 5~km from the nucleus barycentre. In each block all three sensor sides were operated. Measurement times were 100~s or 200~s.

During the tests of the descent with the Philae ground reference model performed at the Deutsches Zentrum f\"ur Luft und Raumfahrt (DLR) in Cologne,  a cross-talk with the Maximum Power Point Tracking (MPPT) electronics of the solar arrays was identified, which resulted in a high rate of false signals interpreted as detected events by the DIM electronics. For the most part, these events were recorded for only a few seconds at the beginning of the blocks. The same behaviour could be observed in the flight data, which means that those measurements in which the number of false signals exceeded 350 could not be used for detecting particle impacts. 

The detection intervals in terms of particle radius for BCT2 measurements during Philae's descent were $[0.5\,\mathrm{mm}-6.5\,\mathrm{mm}]$ and $[0.9\,\mathrm{mm}-6.5\,\mathrm{mm}]$ for margin levels of 40~dB and 50~dB, respectively, used during these measurement blocks \citep{krueger2015}.


\subsubsection{Measurement mode during the first science sequence phase}

\label{subsubsec_fss_op}

In the measurement blocks of the FSS DIM was operated in Burst Continuous (BC) mode. The BC mode delivers the counts for impacts
with a particular $[U_m, T_c]$ combination. The $U_m$ and $T_c$ values are stored in a compressed way in memory cells of different sizes, depending on the expected frequency of such events.
The event times are not registered by the instrument.

Each BC mode measurement lasted for $557-558\,\mathrm{s}$; margins were set either to 30~dB with radius detection interval $[0.25\,\mathrm{mm}-1.5\,\mathrm{mm}]$ or 40~dB with $[0.5\,\mathrm{mm}-1.5\,\mathrm{mm}]$\citep{krueger2015}.


\section{Methods}

\label{sec_methods}


\subsection{Maximum impact rates}

\label{subsec_max_impact_rates}

Provided that only one real impact was registered by DIM during the scientific measurements in the SDL and the FSS phases, it is reasonable to assume an isotropic distribution of the particle trajectories. Moreover, we can suppose that the impacts on the DIM sensor are independent events, hence we can also assume that their occurrence follows a Poisson distribution. For the periods when no detection was made, we seek the detection limit, i.e. the value of the parameter $\lambda$ of the Poisson distribution for which there is an arbitrarily chosen 95\% ($2\sigma$) probability that the number of detected events $N$ will exceed zero in a single measurement

\begin{equation}
   \begin{array}{l}
       P( N > 0 ) = 1- P( N = 0 ) = 1-\dfrac{\lambda^0\exp(-\lambda)}{(0)!}=\\\\
      =1-\exp(-\lambda)=0.95\,, \label{eq_p_limit1a}
   \end{array}
\end{equation} 

thus

\begin{equation}
   \lambda=-\ln(0.05)\approx3\,. \label{eq_p_limit1b}
\end{equation}

For the measurement block, when exactly one real signal was detected, again, only the upper limit of the ambient flux can be estimated. We can define the upper limit of the expected number of impacts as the highest value of $\lambda$ for which there is a 5\% probability that the number of the detected events $N$ will be less than 2 in a single measurement:

\begin{equation}
   \begin{array}{l}
       P( N<2 ) = P( N=0 ) + P( N=1) =\\\\
      = (1+\lambda)\exp(-\lambda)= 0.05\,, \label{eq_p_limit2a}
   \end{array}
\end{equation} 

resulting in 

\begin{equation}
   \lambda\approx4.74\,. \label{eq_p_limit2b}
\end{equation}


\subsection{Geometric factor of the DIM sensor}

\label{subsec_geometric factor}

The relation between the measured impact rates ($N$ in s$^{-1}$) and the particle flux ($\Phi$ in m$^{-2}$s$^{-1}$sr$^{-1}$) in the measurement range of the sensor is given by

\begin{equation}
   N=G\Phi \,, \label{eq_flux}
\end{equation} 

where $G$ is the geometric factor of the detector in m$^2$sr.


\subsubsection{Stand-alone PZT segment}

\label{subsubsec_geom_PZT}

First we consider only one single PZT segment in the XY plane. The geometric factor is given as the sum of the effective areas seen from different directions. For an isotropic particle flux coming from the $Z>0$ half-space, $G_0 = \pi A$, where $A = WL$ is the surface area of one PZT segment (Fig.~\ref{Fig_max_angles}).


\subsubsection{Shading effect of the frame}

\label{subsubsec_geom_frame}

\begin{figure}
   \centering
   \includegraphics[width=\hsize]{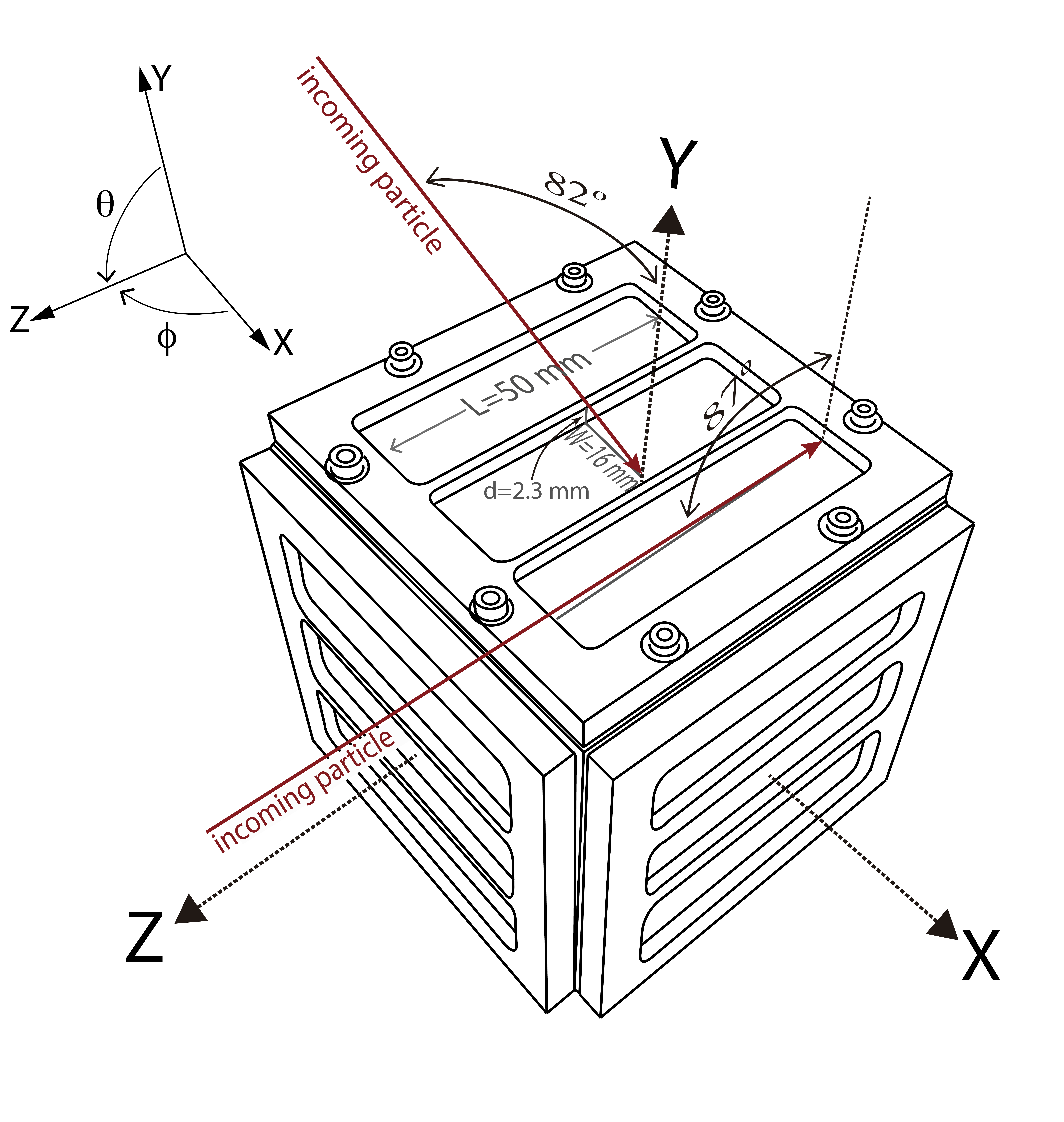}
   \caption{Maximum angles of incidence at which the impacting particles are not shaded by the frame of the sensor.
          }
    \label{Fig_max_angles}
\end{figure}

The outer structure of the DIM sensor that frames the PZTs produces significant shading for particles incident to the sensitive surface under highly oblique angles. The surface of the sensor segments lies 2.3~mm below the outer surface of the frame, which is significant compared to the dimensions of the segments ($W$ = 16~mm; $L$ = 50~mm). For example, if particles come along a direction for which $\phi=0^{\circ}$ and $\theta>87^{\circ}$, or likewise, for which $\phi=90^{\circ}$  and $\theta>82^{\circ}$ the sensor frame completely prevents the particles from reaching the PZT (see Fig.~\ref{Fig_max_angles}).

 The geometric factor, in this case, can be calculated analytically and it is identical to the geometric factor of a radiation detector having rectangular telescope geometry. \citet{thomas1972} derived an analytical formula for the general case of two rectangular areas having sides of length $2X_1$, $2Y_1$ and $2X_2$, $2Y_2$ with $Z$ being their separation. The geometry of one DIM PZT segment with the sensor frame around it corresponds to the special case of equal rectangular areas with dimensions $2X_1 = 2X_2 = L$ and $2Y_1 = 2Y_2 = W$; the separation between the two areas is identical to the depth of the frame $Z = d$. The geometric factor $G_\mathrm{f}$ then can be expressed as

\begin{equation}
   \begin{array}{l}
   G_\mathrm{f}= 2L \sqrt{d^2+W^2}\arctan \dfrac{L}{\sqrt{d^2+W^2}}-2Ld \arctan \dfrac{L}{d}+ \\\\
   +2W \sqrt{d^2+L^2} \arctan \dfrac{W}{\sqrt{d^2+L^2}}-2Wd \arctan \dfrac{W}{d}+ \\\\
   +d^2 \ln \dfrac{(d^2+W^2)(d^2+L^2)}{(d^2+W^2+L^2)d^2}.
   \end{array}
\end{equation}

If we calculate the limit of $G_\mathrm{f}$ as $d$ tends to 0, we get $G_\mathrm{f}=G_{0}$.


\subsubsection{Shading by the structure of Philae}

\label{subsubsec_geom_structure}

In order to consider the shading effects of Philae's structure and the other payloads on the DIM's FoV, numerical simulations were performed with a virtual isotropic particle flux because -- owing to the complexity of the structure -- the problem could not be solved analytically. The DIM sensor was simulated with its frame mount (see Sect.~\ref{subsubsec_geom_frame}), whereas the CAD model of Philae (Bernd Chares, priv. communication) was slightly simplified to reduce computational time, e.g. we neglected the lander feet and reduced the shape of the SD2 drill tower to a properly sized cuboid. The DIM PZT segments were divided into 800 identical squares with surface area of $1\ \mathrm{mm^{2}}$ each. On each square, 32,400 particles were generated with an isotropic flux from a half-space, which means a total number of approximately 78 million particles simulated per DIM side. For each linearly propagating particle the simulation checks if the trail is within DIM's FoV or intersected by Philae's structures. The ratio of actual impacting vs. the total number of simulated particles is named the {detection ratio}.


\subsubsection{Calculated geometric factors}

\label{subsubsec_calc_geom_fact}

The detection ratio for each DIM side and the values of the geometric factor calculated for the cases described in Sect.~\ref{subsec_geometric factor} are summarized in Table \ref{table_geom_factor}. The geometric factors are reduced by 17\% for all three sides if only the shielding effect of the sensor frame is taken into account (inherent shading), whereas if the shielding by the structure and payloads of the lander are also considered the values are reduced by 56\%, 37\%, and 33\% respectively for sensor sides X, Y, and Z.

\begin{table}
\caption{Geometric factors $G$ calculated for the three sensor sides of DIM}             
\label{table_geom_factor}      
\centering          
\begin{tabular}{l c c  c }     
\hline\hline        
Model for the sensor side & \multicolumn{3}{c}{Geometric factor (cm$^2$sr)}\\ 
{} & X & Y & Z \\
\hline                    
   3 stand-alone PZT segments ($3G_0$) & 76.3 & 76.3 & 76.3 \\  
&&&\\[-1.5ex]
   3 PZT segments with frame ($3G_\mathrm{f}$)  & 63.7 & 63.7 & 63.7 \\  
&&&\\[-1.5ex]
   Lander configuration            & 34.0 & 47.9 & 51.0 \\
  \hspace*{10pt} (relative to $3G_0$)	     &  (-56\%) & (-37\%) & (-33\%) \\
  \hspace*{10pt} (detection ratio)               & (0.445) & (0.628) & (0.668) \\
\hline                  
\end{tabular}
\end{table}

The detection ratio decreases to the corners and the edges for each segment owing to the shading by the DIM frame. The least shielding occurs for the Z-side of DIM. This side is only slightly shielded by the SD2 drill tower and the DIM mounting frames. The analytical calculations performed in Sect.~\ref{subsubsec_geom_frame} also served for verification of the numerical model used in Sect.~\ref{subsubsec_geom_structure}.


\section{Results}

\label{sec_results}
 

\subsection{Estimation of the maximum flux during descent}

\label{subsec_flux_descent}

Estimations for the upper limit of the flux during descent were performed only for those measurements for which the effective measurement time was higher than 0~s. The results are summarized in Table~\ref{table_flux_sdl}. The sequential number of the measurements in Col.~1 are according to \citet{krueger2015}. Column~9 corresponds only to the number of real detections; false signals are not included. The maximum impact rate ($N_{\mathrm{max.}}$) in Col.~10 was defined as the ratio of parameter $\lambda$ of the Poisson distribution calculated according to Eqs.~(\ref{eq_p_limit1b}) and (\ref{eq_p_limit2b}) respectively for the non-detection and the single detection of particle impacts to the effective measurement time shown in Col.~6. Finally, $\Phi_{\mathrm{max.}}$ in Col.~11 was calculated according to Eq.~(\ref{eq_flux}).

\begin{table*}
\caption{Upper limit of the ambient particle flux during descent (12 Nov 2014)}             
\label{table_flux_sdl}      
\centering          
\begin{tabular}{c c c c c c c c c c c}     
\hline\hline        
Meas. & Start   &   Dist.  &  DIM  &  Meas. &  Eff. meas.  &  Margin  &  Meas.    & Det. &   
$N_{\mathrm{max.}}$   & $\Phi_{\mathrm{max.}}$ \\
numb. & time    &     {}    &  side  &  time   &      time      &    {}       &  range   &    events  &   {}      &     {}           \\         
{}  & (UTC)  &   (km)  &   {}     &   (s)    &       (s)        &     (dB)   &     (mm)        &     {}        &  ($\mathrm{s}^{-1}$) & ($\mathrm{m}^{-2}\mathrm{s}^{-1}\mathrm{sr}^{-1})$      \\ 
(1) &  (2) &   (3)  & (4) & (5) & (6) & (7) & (8) & (9) & (10) & (11)  \\
\hline                    
30 & 08:38:32 & 22.2 & X & 200 & 0 & 40 & 0.5 - 6.5 & 0 & -- & -- \\
31 & 08:42:23 & 22.1 & Y & 200 & 0 & 40 & 0.5 - 6.5 & 0 & -- & -- \\
32 & 08:46:13 & 22.0 & Z & 200 & 198 & 40 & 0.5 - 6.5 & 0 & $1.5\cdot10^{-2}$ & $3\cdot10^{-8}$ \\
&&&&&&&&&&\\[-1.5ex]
33 & 08:50:03 & 21.8 & X & 200 & 200 & 50 & 0.9 - 6.5 & 0 & $1.5\cdot10^{-2}$ & $4\cdot10^{-8}$  \\
34 & 08:53:52 & 21.6 & Y & 200 & 200 & 50 & 0.9 - 6.5 & 0 & $1.5\cdot10^{-2}$ & $3\cdot10^{-8}$  \\
35 & 08:57:42 & 21.4 & Z & 200 & 200 & 50 & 0.9 - 6.5 & 0 & $1.5\cdot10^{-2}$ & $3\cdot10^{-8}$  \\
&&&&&&&&&&\\[-1.5ex]
36 & 09:59:04 & 18.6 & X & 200 & 0 & 40 & 0.5 - 6.5 & 0 & -- & -- \\
37 & 10:02:54 & 18.4 & Y & 200 & 198 & 40 & 0.5 - 6.5 & 0 & $1.5\cdot10^{-2}$ & $3\cdot10^{-8}$  \\
38 & 10:06:44 & 18.3 & Z & 200 & 0 & 40 & 0.5 - 6.5 & 0 & -- & -- \\
&&&&&&&&&&\\[-1.5ex]
39 & 14:40:04 &   5.1 & X & 100 &   96 & 40 & 0.5 - 6.5 & 0 & $3.1\cdot10^{-2}$ & $9\cdot10^{-8}$ \\
40 & 14:42:14 &   5.0 & Y & 100 &   98 & 40 & 0.5 - 6.5 & 1 & $4.8\cdot10^{-2}$ & $1\cdot10^{-7}$ \\
41 & 14:44:24 &   4.9 & Z & 100 &     0 & 40 & 0.5 - 6.5 & 0 & -- & -- \\
\hline                  
\end{tabular}
\end{table*}


\subsection{Estimation of the maximum flux at Abydos}

\label{subsec_flux_abydos}

Since no impact signals were detected at all at the final landing site Abydos during the FSS,
the measurement times for each sensor side were summed:

\begin{equation}
  T_{\mathrm{meas.,}\,i,\,\mathrm{total}}= \sum\limits_{j=1}^{5}\sum\limits_{k=1}^{2} T_{\mathrm{meas.,}\,i,\,j,\,k},
\end{equation}

where $T_\mathrm{meas.,\,i,\,j,\,k}$ is the measurement time for sensor side $i$ in the $k^{th}$ measurement of FSS DIM Block number $j$.  The maximum impact rate for the $i^{th}$ sensor side in the above case is

\begin{equation}
   N_{\mathrm{imp.},\, i,\,\mathrm{max.}}= \lambda_\mathrm{max.} /  T_{\mathrm{meas.,}\,i,\,\mathrm{total}}\,,
\end{equation}

where $\lambda_\mathrm{max.}$ is calculated according to Eq. \ref{eq_p_limit1b}.

The upper limits of the flux values ($\Phi_{\mathrm{max.}}$, see Table \ref{table_flux_fss}) for particles in the measurement range of the DIM sensor were estimated using Eq. \ref{eq_flux} and by considering the geometric factors calculated for the lander configuration (see values in Table \ref{table_geom_factor}).

\begin{table}
\caption{
Particle flux calculated for the three sensor sides of DIM after landing at Abydos}             
\label{table_flux_fss}      
\centering          
\begin{tabular}{c c c c }     
\hline\hline        
Sensor side & Meas. time & $N_{\mathrm{max.}}$ & $\Phi_{\mathrm{max.}}$ \\ 
{} & (s) & ($\mathrm{s}^{-1}$) & ($\mathrm{m}^{-2}\mathrm{s}^{-1}\mathrm{sr}^{-1})$ \\
\hline                    
   +X      & 5579 & $5.4\cdot 10^{-4}$ & $1.6\cdot 10^{-9}$ \\  
   +Y      & 5579 & $5.4\cdot 10^{-4}$ & $1.1\cdot 10^{-9}$ \\  
   +Z      & 5579 & $5.4\cdot 10^{-4}$ & $1.1\cdot 10^{-9}$ \\  
\hline                  
\end{tabular}
\tablefoot{The measurement times of the individual measurement blocks during FSS were reported by \citet{krueger2015}.} 
\end{table}

\subsection{Particle flux and the topography of Abydos}
\label{subsec_flux_topography_abydos}

\begin{figure}
   \centering
   \includegraphics[width=\hsize]{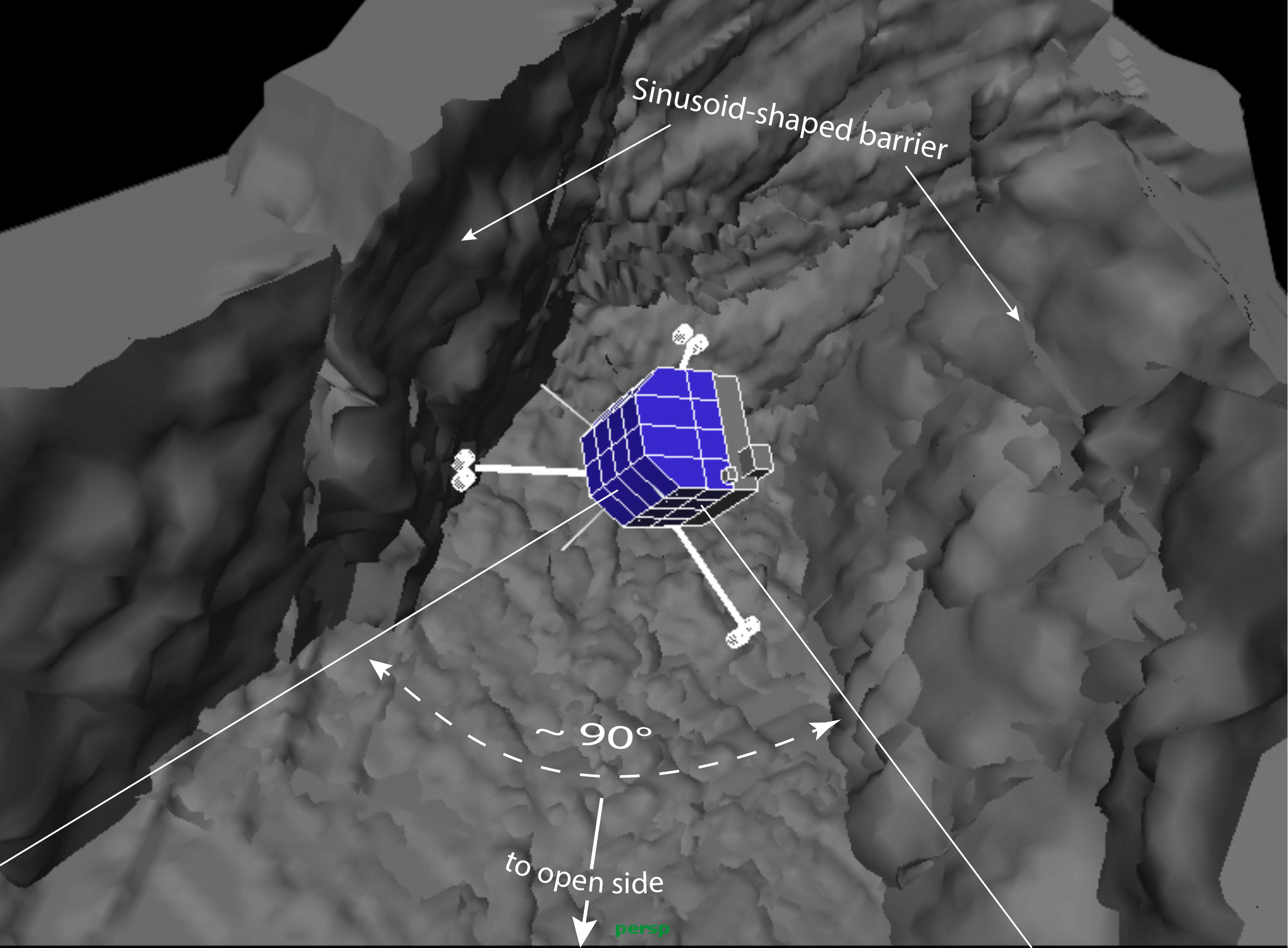}
   \caption{Diagram of the basic geometry used in the simulation explained in Sect.~\ref{subsec_flux_topography_abydos}. The lander Philae is surrounded by a barrier curved as a sinusoid. The front (bottom of the image) and the space above the lander are open. The illumination of the Sun mainly comes from the front because the Sun moves on a plane that forms $<22^{\circ}$ with the floor of this box. The $\approx{90^{\circ}}$ angle indicates the width of the field of view of the lander towards the Sun.
          }
    \label{Fig_ABYDOS_DIAG}
\end{figure}

The topography at Abydos is a major concern with regard to the potential flux detection, in addition to the activity of the comet and the self shading by the structure of Philae or the sensor itself. The images of the  \c{C}iva cameras \citep{bibring2015} at Abydos show that the lander is partially surrounded by an obstacle that prevents the sunlight from reaching the solar panels for long periods of time. Right after landing, in November 2014, the panels received sunlight for less than two hours per comet day, which was not enough to charge the lander's batteries. Because understanding the illumination conditions on the lander would help us to determine the topography at Abydos, Flandes et al. (2016; document in preparation) simulated the pattern of illumination on Philae assuming that the lander is partially surrounded by a sinusoid-shaped barrier with a height equivalent to approximately three times the height of the lander (see Fig. \ref{Fig_ABYDOS_DIAG}). A top view of this configuration would set the lander in a trough or peak of this sinusoid. In this simulation the Sun always moves along the horizon for the lander (at low elevations $<22.5^{\circ}$). Under these conditions, the lander is illuminated for 1.5 hours out of the 12.4 hour rotation period of the comet. The panel that receives the largest amount of light is Panel 1 (which is parallel with the Y side of DIM), followed by Panel 2. Panel 6 (top panel, which is parallel with the Z side of DIM) and Panel 3 receive very little sunlight. 

If this geometry is accurate, the incoming particle flux for DIM will be very much reduced as well (assuming that neither the barrier nor the floor of Abydos in the immediate vicinity of the spacecraft are dust sources). The temperature measured by the Multi-Purpose Sensors for Surface and Subsurface Science (MUPUS) experiment at 90--130~K is well below the water sublimation temperature, which implies that the immediate environment was not active \citep{spohn2015}. For practical purposes, the lander could be considered to be inside a hypothetical partially opened box where the floor of the site is its bottom and the barrier/wall represents only three of its sides. One conclusion is that particles could reach the sensor only through the top of this box or from one of its sides. With this further simplification, the general field of view of the sensor (if considered at the centre of this box) would be reduced to $1/3$ of the total, i.e. $4/3\pi\,\mathrm{sr}$. 
Given that the Sun never goes far above the horizon in this simulation, the three most important parameters are the separation between the sides, the depth of this box, and the maximum elevation of the Sun. A variation of $>10\%$ in any of these three parameters would produce a similar variation in the illumination pattern on the lander. Still, the geometric factor of the sensor segments is already reduced to 33\%--56\% as a result of shading by the lander structure and some other payloads (see Table~\ref{table_geom_factor}), which means that further corrections are unnecessary as the uncertainties in other factors are significantly higher.

\citet{auster2015} estimated the local gravity as $g\,=\,10^{-4}\,\mathrm{m~s}^{-2}$ at the Abydos landing site at a distance of 2332~m from the barycenter of the nucleus. This gives an escape velocity of approximately $0.7\,\mathrm{m~s}^{-1}$. If particle speeds $v$ below the escape velocity are considered, the expected speed range for particles in the measurement range of DIM can be constrained to $0.1\,\mathrm{m~s}^{-1}-0.7\,\mathrm{m~s}^{-1}$. An upper boundary for the volume density ($n_{\mathrm{max.}} = \Phi_{\mathrm{max.}}\Omega_{\mathrm{eff.}}/v$, where $\Omega_{\mathrm{eff.}}$ is the effective solid angle) of such particles is then of the order of $10^{-11}\,\mathrm{m}^{-3}-10^{-12}\,\mathrm{m}^{-3}$.


\section{Discussion}

\label{sec_discussion}
 
Although DIM was taking measurements starting from the onset of comet activity of 67P/C-G in November 2014 at about 3~AU heliocentric distance, only one single detection of particle impact was made. During descent the relatively short measurement times did not permit the measurement of particles with good statistics. The total measurement time was only 35~minutes, which -- owing to the large number of false signals resulting from the cross-talk from the MPPTs -- was further reduced to less than 20~minutes noise-free time. Hence, instead of the flux, the upper limit of the flux of particles in the measurement range of DIM was determined (see Table~\ref{table_flux_sdl}).

At the final landing site Abydos, the total measurement time was an order of magnitude higher than for the descent, but -- probably due to the shading from the environment and the low activity of the comet in the vicinity of Philae -- no detection of particle impacts was made.

Calculations were also performed with the GIADA Performance Simulator (GIPSI) to simulate the expected fluxes on the DIM instrument during the descent of the lander. GIADA is an experiment on the Rosetta orbiter devoted to the measurements of the physical properties of cometary dust grains in orbit around the nucleus. The grain detection system (GDS) of GIADA detects the transit of each single grain entering the instrument by means of a light curtain. In addition an impact sensor (IS) equipped with PZT sensors and five microbalances measuring mass deposition on quartz crystals are included in the experiment \citep{colangeli2007}. The Java client software GIPSI is able to simulate GIADA performance, in particular GIPSI forecasts how an instrument or a defined surface reacts to a dust environment along a specific trajectory for defined time intervals. Inputs to GIPSI are the dust environments described or evaluated by models (e.g. grain number density, particle size distribution, and velocity). As inputs to GIPSI, in addition to the time-dependent 3D model environment, we use the spacecraft and comet orbits (the attitude and the position of the spacecraft and the speed along the orbit for each time step) and instrument parameters (i.e. field of view, sensitive surface area, subsystems sensitivities). GIPSI simulates the instrument-dust interaction for each time step (defined as input of the simulation) and position along the orbit of the spacecraft. The software evaluates the vector parameters of the dust model outputs into the instrument reference frame considering the rotation of the comet nucleus and the velocity of the spacecraft. The outputs of the simulation are the number of grains impacting on the surface of the sensor \citep{dellacorte2014}.

In order to simulate the expected fluxes on the DIM instrument during the descent of the lander we used the GIPSI simulation software considering as dust environment the fluxes and the speeds described in \citet{fulle2010}. The dynamical parameters measured by GIADA during the same period confirm that the use of Fulle's model is substantiated \citep{dellacorte2015, rotundi2015}. Fulle’s dust-tail model \citep{fulle1987, fulle1989} derives an ejection flux and an ejection velocity for each dust mass \citep{fulle1992}. The grains are assumed to be spherical. According to the derived dust ejection velocity, the mass able to escape the nucleus gravity field at a distance of 20 nucleus radii, assuming a bulk density of $1000\,\mathrm{kg}~\mathrm{m}^{-3}$, is checked. Owing to the nucleus asphericity and possibly lower bulk density ($533\,\mathrm{kg}~\mathrm{m}^{-3}$ as reported by \citet{paetzold2016}), the escape velocity is probably significantly lower than the assumed value of $0.5~\mathrm{m~s}^{-1}$ \citep{fulle2010}. The method also requires the value of grain-specific mass to be postulated. The lowest values of the dust mass loss rate is between 10 and $40~\mathrm{kg~s}^{-1}$ at 3~AU, derived by the assumed specific mass and the reported maximum dust mass loss rate $112~\mathrm{kg~s}^{-1}$ at 3~AU.  The model assigns to all grains of the same size bin a constant radial velocity equal to the terminal velocity computed in \citet{fulle2010}.
In particular, we used the upper values for the fluxes reported in \citet{fulle2010}. We considered an isotropic expansion of the particle flux, and to calculate the number of particles emitted, we considered two different densities for the particles; $100\,\mathrm{kg}~\mathrm{m}^{-3}$ (fluffy) and $1000\,\mathrm{kg}~\mathrm{m}^{-3}$ (compact). Fluffy particles emitted from the comet nucleus were detected in the size range up to a few hundred micrometres by the Cometary Secondary Ion Mass Analyzer (COSIMA) on board the Rosetta orbiter \citep{schulz2015, langevin2016}. From the GDS-only detections of the GIADA experiment, \citet{fulle2015} also inferred fluffy particles of equivalent bulk density of less than $1\,\mathrm{kg}~\mathrm{m}^{-3}$, which they associated with the fluffy particles observed by COSIMA.

 For the trajectory of Philae we used the spice kernel reported in the ESAC repository (LORL\_DL\_007\_02\_\_\_\_P\_\_00220.BSP). Since the GIPSI tool is only able to simulate the fluxes over the Rosetta Orbiter spacecraft, to simulate the fluxes over the DIM surface we modified the kernel by changing the reference object of the trajectory and imposing the lander as reference object. Owing to the lack of a consolidated spice kernel describing the lander attitude during the descent we considered an orientation with the +Z side of the lander parallel to the direction of the force of gravity (nadir direction) at the given position.

\begin{figure}
   \centering
   \includegraphics[width=\hsize]{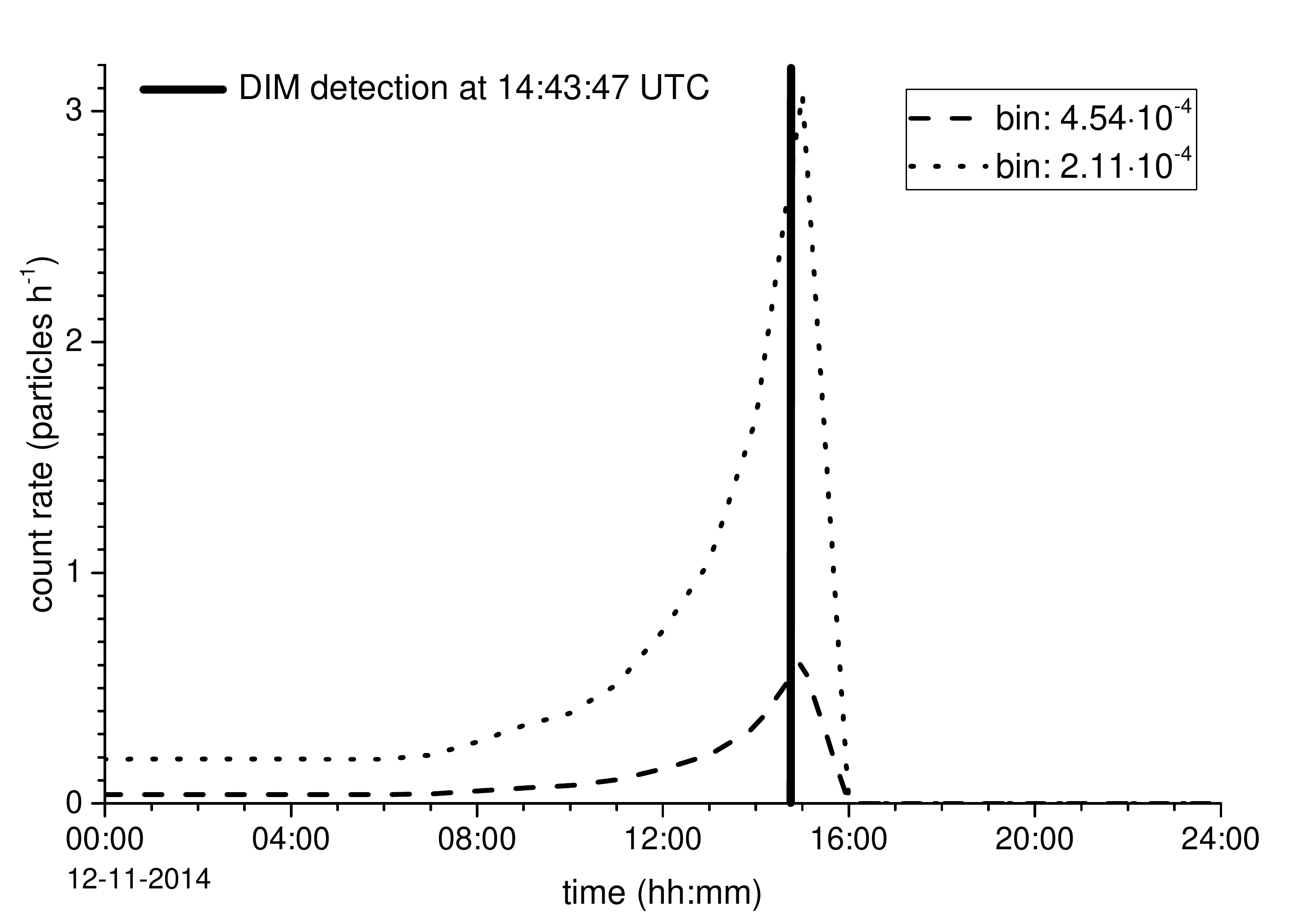}
   \caption{Count rates of compact particles in the size bins with radius 0.21~mm and 0.45~mm as calculated with GIPSI for a collecting surface area of $100\,\mathrm{cm}^{2}$ and a field of view of 67$^{\circ}$ (GDS configuration).}
    \label{Fig_GIADA_compact}
\end{figure}

\begin{figure}
   \centering
   \includegraphics[width=\hsize]{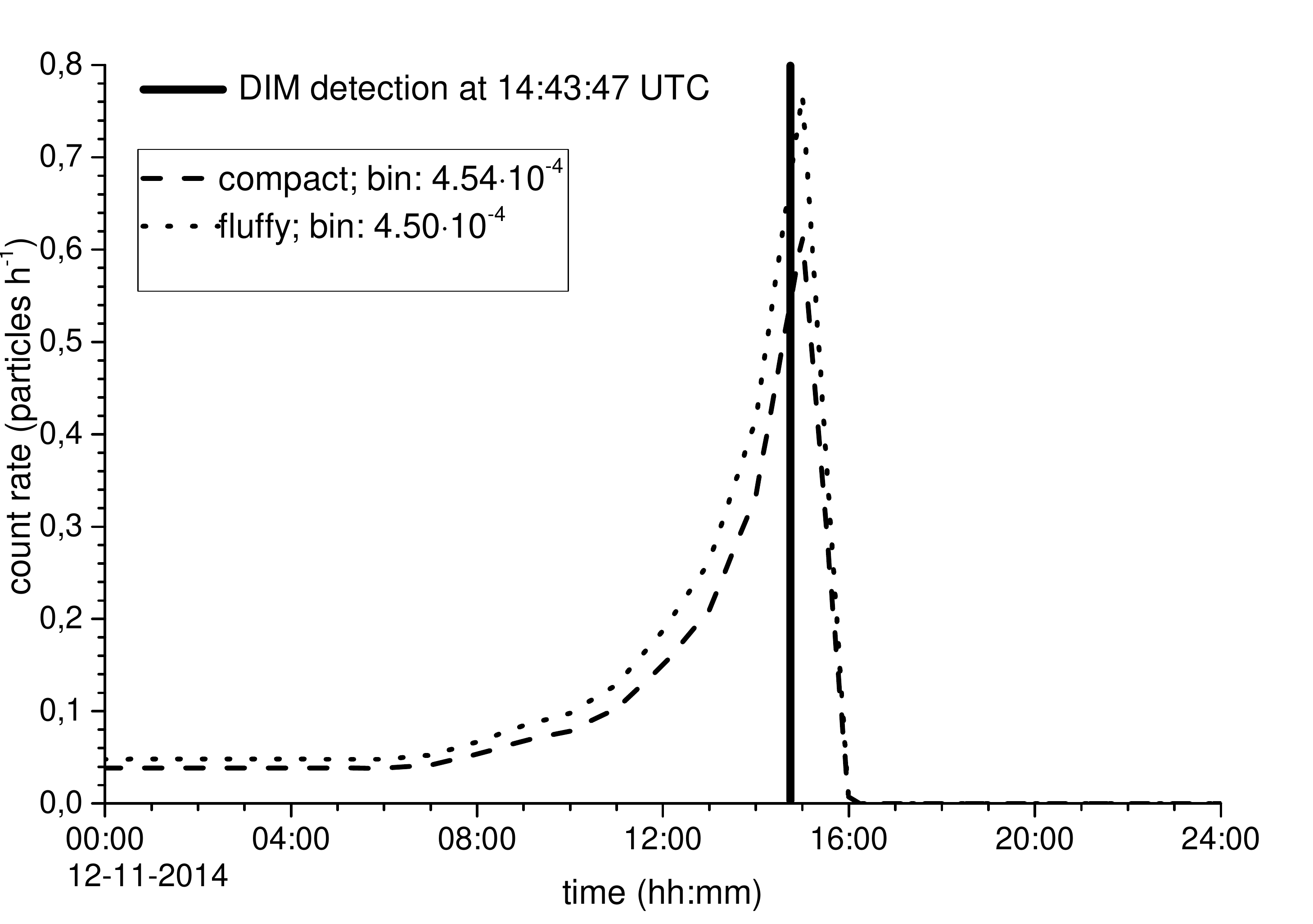}
   \caption{Count rates of compact and fluffy particles in the size bin of particles with radius 0.45~mm as calculated with GIPSI for a collecting surface area of $100\,\mathrm{cm}^{2}$ and a field of view of 67$^{\circ}$ (GDS configuration).}
    \label{Fig_GIADA_0_5mm_comp}
\end{figure}

The DIM sensor is sensitive to compact particles having a minimum radius of 0.25~mm and 0.5~mm if the detection margin is set to 30~dB and 40~dB, respectively. The particles have a power law size distribution, so it is enough to consider the size bins of particles with radius of 0.21~mm and 0.45~mm in the first approximation. The results of the simulation for compact particles are shown in Fig.~\ref{Fig_GIADA_compact}. For fluffy particles the corresponding count rates are only 25\% higher (for comparison of data see Fig.~\ref{Fig_GIADA_0_5mm_comp}). The orientation of Philae during DIM operational periods was not known at the time of writing, hence GIPSI fluxes calculated in the nadir direction are used only as the upper limit. In the GIPSI tool the configuration of the GDS in GIADA, having a field of view of 67$^{\circ}$ (corresponding to 1~sr) and a collecting surface of $100\,\mathrm{cm}^{2}$ are considered. This approximation results in an expected number of counts of maximum $0.005-0.5$ for compact and for fluffy particles within a total measurement time of 20 minutes. This is in good agreement with the non-detection of compact particles during descent. Nevertheless, the detection of a fluffy particle with a radius of 1~mm was also an extremely unlikely event. Since the sensor sides of DIM were pointing off-nadir, a GIPSI calculation knowing the orientiation of the lander would have provided even lower values for the number of counts.

\section{Conclusions}

\label{sec_conclusions}
 
On the 12 November 2014 lander Philae was deployed from the Rosetta Orbiter onto the nucleus of comet 67P/C-G. The DIM experiment of the SESAME instrument package was switched on several times during descent and on the surface of the nucleus to detect impacts from submillimetre- to millimetre-sized particles.

Based on measurements performed with DIM, the upper limit of the flux of particles in the measurement range of the instrument was of the order of $10^{-8}-10^{-7}\mathrm{m}^{-2}\mathrm{s}^{-1}\mathrm{sr}^{-1}$ during descent. The upper limit of the ambient flux of the submillimetre- and millimetre-sized dust and ice particles at Abydos was estimated to be $1.6\cdot10^{-9} \mathrm{m}^{-2}\mathrm{s}^{-1}\mathrm{sr}^{-1}$ on 13 and 14 November 2014. A correction factor of $1/3$ for the field of view of the sensors  was calculated based on an analysis of the pattern of illumination on Philae. Considering particle speeds below escape velocity, the upper limit for the volume density of particles in the measurement range of DIM was constrained to $10^{-11}\,\mathrm{m}^{-3}-10^{-12}\,\mathrm{m}^{-3}$. 
 
Results of the calculations performed with the GIPSI tool on the expected particle fluxes during the descent of Philae are compatible with the non-detection of compact particles by the DIM instrument.



\begin{acknowledgements}

SESAME is an experiment on the Rosetta lander Philae. It consists of three instruments CASSE, DIM, and PP, which were provided by a consortium comprising DLR, MPS, FMI, MTA EK, Fraunhofer IZFP, Univ. Cologne, LATMOS, and ESTEC. The contribution from MTA EK to the SESAME-DIM experiment was co-funded through the PRODEX contract No. 90010 and by the Government of Hungary through European Space Agency contracts No. 98001, 98072, 4000106879/12/NL/KML, and 4000107211/12/NL/KML under the plan for European Cooperating States (PECS). This research was supported by the German Bundesministerium f\"ur Bildung und Forschung through Deutsches Zentrum f\"ur Luft- und Raumfahrt e.V. (DLR, grant 50 QP 1302). The work published in this paper was performed during a visit of A. Hirn and A. Flandes at MPS. Both guest scientists are grateful to MPS for financial support during the visit. A. Flandes was also supported by DGAPA-PAPIIT IA100114 and IA105016. We thank the Rosetta project at ESA and the Philae project at DLR and CNES for effective and successful mission operations.
 
\end{acknowledgements}

   \bibliographystyle{aa} 
   \bibliography{AA_DIMflux_submitted_final_18_may_2016} 

\end{document}